\documentclass[11pt]{article}
\usepackage{hyperref}
\pdfoutput=1
\title{Stability of a jet in crossflow}

\author{Milo\v s Ilak, Philipp Schlatter, Shervin Bagheri, \\
             Mattias Chevalier and Dan S. Henningson\\
             \\Linn\' e FLOW Centre\\
             and\\
             Swedish e-Science Research Centre (SeRC)\\
             KTH Mechanics, SE-100 44 Stockholm, Sweden}

\date{}

\begin{document}

\maketitle

\abstract{We have produced a fluid dynamics video with data from Direct Numerical Simulation (DNS) of a jet in crossflow at several low values of the velocity inflow ratio $R$. We show that, as the velocity ratio $R$ increases, the flow evolves from simple periodic vortex shedding (a limit cycle) to more complicated quasi-periodic behavior, before finally exhibiting asymmetric chaotic motion. We also perform a stability analysis just above the first bifurcation, where $R$ is the bifurcation parameter. Using the overlap of the direct and the adjoint eigenmodes, we confirm that the first instability arises in the shear layer downstream of the jet orifice on the boundary of the backflow region just behind the jet.}

\paragraph{Introduction}

An understanding of the jet in crossflow is crucial to many practical applications, and this flow has been the subject of a large number of experimental and numerical studies over the last several decades. An overview of the major results and current efforts may be found for example in~\cite{Karagozian10} and the references therein. This contribution focuses on the stability of the incompressible, non-reactive jet in crossflow, which we study through numerical simulations. 

The videos included show the Direct Numerical Simulation (DNS) of a jet in crossflow at different values of the ratio of the jet exit velocity and the boundary layer free-stream velocity, defined as $R$, which is the bifurcation parameter in the current investigation. The lowest $R$ shown is $R=0.7$, which is just above the first bifurcation, and the highest $R$ shown is $R=3$, which is the simulation of~\cite{Bagheri-09}. We show that, as $R$ is increased, the flow gradually evolves from simple shedding of hairpin vortices downstream of the jet, through a more complex shedding pattern at $R=2$, which is still symmetric, to the turbulent state observed at $R=3$. In order to characterize the first instability, we perform a stability analysis at $R=0.7$. We show the leading unstable eigenmode, and the corresponding adjoint eigenmode, and their overlap, which is known as a `wavemaker'~\cite{Chomaz-05,Giannetti-07}. The wavemaker region is found to be in the region of high shear on the boundary of the backflow region and the injected fluid just downstream of the jet orifice. A part of the video shows the oscillating shear layer, illustrated by the spanwise vorticity. Furthermore, we show that at a higher value of $R$, $R=0.8$, even though multiple unstable eigenmodes are present, the flow still saturates to a simple limit cycle whose frequency is the frequency of the most unstable eigenmode.

\paragraph{Direct Numerical Simulation}
The fully spectral SIMSON code was used for the simulations~\cite{SIMSON-manual}. The resolution of all computations was $256 \times 201 \times 144$, and the box size is $75.0 \times 20.0 \times 30.0$ in the streamwise, wall-normal and spanwise directions respectively. The jet is imposed as a Dirichlet boundary condition on the wall.

The jet in crossflow is characterized by three dimensionless parameters - the Reynolds number based on the free-stream velocity $U_{\infty}$ of the boundary layer and the boundary layer displacement thickness $\delta_0^*$ at domain entrance, $Re=U_{\infty}\delta_0^*/\nu$, where $\nu$ is the kinematic viscosity, the jet Reynolds number based on the jet velocity and the jet diamater, $Re_{jet}=V_jD/\nu$, and the ratio of the jet velocity to the free-stream velocity, $R=V_j/U_{\infty}$. In this work $Re=165$ and the jet diameter is set to be $D=3\delta_0^*$, so that $R$ is the only parameter that changes. The value of $V_j$ used for the definition of the velocity ratio $R$ varies in literature. While the value used most often is the bulk or mean exit velocity of the jet at the orifice, here we adopt the definition of~\cite{Bagheri-09} and use the jet centerline exit velocity. All runs were performed using up to 256 parallel processors.

\paragraph{Stability analysis}
The direct and adjoint eigenmodes were computed via the Implicitly Restarted Arnoldi Method (IRAM)~\cite{arpack98} using a linearized SIMSON time-stepper with the same box size and resolution as the DNS runs. The steady-state solutions of the Navier-Stokes equations (base flows) required for the linearization were computed using Selective Frequency Damping (SFD)~\cite{AakervikBHHMS-06}. More details about the simulation setup and the use of IRAM for stability analysis in this setup can be found in~\cite{Bagheri-09}. 

\paragraph{Visualization}
The visualization was done using ParaView scripted through Python for greater flexibility in creating the animations. Most visualizations in the video show volume rendering of the $\lambda_2$ vortex identification criterion~\cite{JeongH-95}, with `vortex cores', i.e., regions of highly negative values of $\lambda_2$, shown in yellow and `vortex edges', or regions with $\lambda_2$ negative but closer to zero, shown in light brown. The visualization that illustrates that the location of the `wavemaker' is in the backflow region was done using isocontour plots. 

\paragraph{Acknowledgments}

Computer time was provided by the Swedish National Infrastructure for Computing (SNIC) with a generous grant by the Knut and Alice Wallenberg Foundation (KAW). We would also like to thank Johan Raber from NSC and Jonathan Vincent from PDC for their help with the visualization, and Jan Pralits, Simone Camarri, Luca Brandt and Outi Tammisola for helpful discussions.



\end{document}